\documentclass[aps,prl,twocolumn,double-spaced,superscriptaddress]{revtex4}

\usepackage{graphicx}
\usepackage{amssymb}

\begin{document}


\title{Angular dependence of the melting of the porous vortex matter in $\mathrm{Bi_{2}Sr_{2}CaCu_{2}O_{8}}$}


\author{Nurit Avraham}
\affiliation{Department of Condensed Matter Physics, The Weizmann
Institute of Science, Rehovot 76100, Israel}
\author{Y. Y. Goldschmidt}
\author{J. T. Liu }
\affiliation{Department of Physics and Astronomy, University of
Pittsburgh, Pittsburgh, Pennsylvania 15260, USA}
\author{Y. Myasoedov}
\author{M. Rappaport}
\author{E. Zeldov}
\affiliation{Department of Condensed Matter Physics, The Weizmann
Institute of Science, Rehovot 76100, Israel}
\author{C. J. van der Beek}
\author{M. Konczykowski}
\affiliation{Laboratoire des Solides Irradie´s, CNRS UMR 7642 and
CEA-CMS-DRECAM, Ecole Polytechnique, 91128 Palaiseau, France}
\author{T. Tamegai}
\affiliation{ Department of Applied Physics, The
University of Tokyo, Hongo, Bunkyo-ku, Tokyo 113-8656, Japan}

\date{\today}

\begin{abstract}

Vortex matter in $\mathrm{Bi_{2}Sr_{2}CaCu_{2}O_{8}}$ with a low
concentration of tilted columnar defects (CDs) was studied using
magneto-optical measurements and molecular dynamics simulations. It is
found that while the dynamic properties are significantly affected by
tilting the magnetic field away from the CDs, the thermodynamic
transitions are angle independent. The simulations indicate that vortex
pancakes remain localized on the CDs even at large tilting angles.
This preserves the vortex thermodynamics, while vortex pinning is
considerably weakened  due to  kink sliding.
\end{abstract}


\maketitle

The dynamic and thermodynamic properties of vortex matter in high
$T_{c}$ superconductors are strongly affected by the introduction of
correlated disorder in the form of columnar defects (CDs). In the
presence of a low concentration of CDs, experimental \cite{banerjee,
banerjgold, menghini} and theoretical results \cite{radzihovsky,
tyagigold, dasgupta, nonomura, lopatin, goldcuan} indicate the
presence of two vortex subsystems: vortices localized on pinning
centers and interstitial vortices between them. At low temperature,
a \emph{porous vortex solid} phase is created. A rigid matrix of
vortices, strongly pinned on a network of random CDs, is formed,
while the interstitial vortices create relatively ordered
nanocrystals within the pores of the matrix. When the temperature is
increased, the interstitial vortices melt while the rigid matrix
remains localized, forming a \emph{nanoliquid} phase in which
nanodroplets of liquid are weakly confined within the pores of the
matrix. When the temperature is increased further, the matrix
delocalizes at the delocalization transition, and the liquid becomes
\emph{homogeneous}. Above this transition the effect of the CDs on
vortex dynamics becomes indiscernible.

This picture holds when the angle $\theta$ between the average
induction $\mathbf{B}$ and the CDs is zero. According to theoretical
models \cite{nelson,blatter}, for small $\theta$, the localized
vortices on CDs remain completely trapped along the CDs. At larger
$\theta$, the vortices form a staircase structure with segments
trapped on different CDs connected by weakly pinned kinks. At still
larger angles, the vortices follow the field direction and are not
affected by the correlated nature of the CDs. In general, one
expects the effect of the correlated nature of the CDs to decrease
as $\theta$ increases, and to vanish at $\theta=90^{\circ}$. In more
isotropic materials such as $\mathrm{YBa_{2}Cu_{3}O_{7}}$, this
picture is supported by various experiments, showing an angle
dependent behavior in both dynamic \cite{civale} and thermodynamic
\cite{hayani} properties. In layered materials such as
$\mathrm{Bi_{2}Sr_{2}CaCu_{2}O_{8}}$ (BSCCO), vortices consist of
stacks of two-dimensional pancake vortices (PVs) defined in the
superconducting CuO$_{2}$ layers only. Measurements of the
equilibrium torque and magnetization in BSCCO \cite{drost} have
shown that the CD occupation by PVs in the liquid phase is angle
independent. In contrast, Josephson plasma resonance measurements
\cite{Tamegay} show that the PV alignment in the liquid is
significantly affected by CDs. The corresponding equilibrium
properties of the solid phase have not been studied. Since the
melting transition depends on the difference in the behavior of the
free energies of the solid and liquid, it is very interesting to
study the angular dependence of the melting in the presence of CDs.
Moreover, CDs are known to shift the first-order melting line upward
\cite{banerjee} and point disorder was found to shift it downward
\cite{boris}. One might expect, therefore, that tilting $\mathbf{B}$
away from the CDs would shift the melting line continuously from
above the pristine melting line to below it, reaching the lowest
position at $\theta=90^{\circ}$, where CDs act like point defects.

Here, we study the angular dependence of the dynamic and
thermodynamic properties of porous vortex matter in BSCCO. Our
measurements show that the irreversibility line shifts to lower $T$
when the field is tilted away from the CDs, as expected
\cite{civale,irreversible}. However, the thermodynamic melting and
delocalization lines have \em no \rm angular dependence. Our
simulations show that even at large angles the PVs are effectively
pinned by the CDs and therefore the thermodynamic features remain
unaffected. On the other hand, the formation of a kink structure of
the PV stacks increases their mobility due to kink sliding,
therefore suppressing the irreversible properties.

For the measurements, BSCCO crystals ($T_{c} \approx 90$ K) were
irradiated, at GANIL, with 1 GeV Pb ions, through stainless steel
masks with triangular arrays of 90 $\mu$m holes. This produces
amorphous columnar tracks, with density $n_d$ equal to the ion dose,
only in those parts of the crystals situated under the holes.
Crystals A and B, irradiated at 45$^{\circ}$ with respect to the
$c$-axis, have matching fields $B_{\phi} = n_d \Phi_{0}$ of 20 G and
30 G respectively. The measurements were performed using
differential magneto-optical imaging (DMO) \cite{soibel} with a
vector magnet that allows application of a DC magnetic field in any
direction, so that the local magnetic response can be imaged over
the full range of angles from ${\bf B} \parallel $ CDs ($\theta =
0$) to ${\bf B} \perp $ CDs ($\theta = 90^{\circ}$).

\begin{figure}
\includegraphics[width=0.40\textwidth]{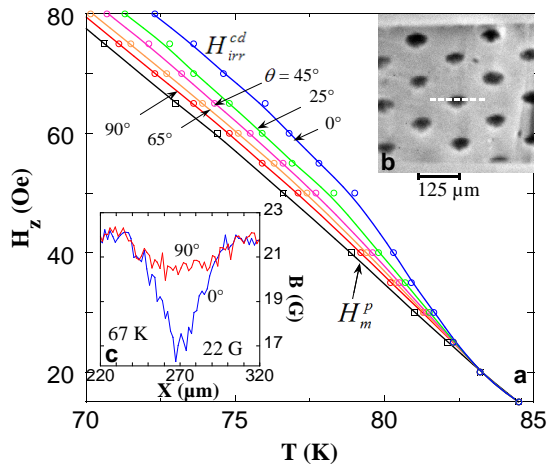}
\caption{\label{fig1} (color online)
(a) The irreversibility line $H_{irr}^{cd}(T)$ of crystal A, mapped
for five angles $\theta$ between $\mathbf{B}$ and the CDs, from
$\theta=0$ to $90^{\circ}$. $H_{m}^{p}$ is the pristine melting
line. (b) DMO image of part of the sample at $T=79$ K and $H_{z}=35$
Oe. (c) Flux profiles across the irradiated region [dashed line in
(b)], at 67 K and $H_{z}=22$ Oe, for $\theta = 0^{\circ}$ and
$90^{\circ}$.}
\end{figure}

Figure 1(a) shows  the irreversibility line of the irradiated
regions $H_{irr}^{cd}(T)$ on an $H_{z}-T$ plane, where $H_{z}$ is
the c-axis component of the applied field, for different angles of
$\mathbf{B}$ with respect to the CDs \cite{angle}. The curves were
obtained using DMO with field modulation of 1 Oe, upon sweeping the
temperature. Below $H_{irr}^{cd}(T)$, the external field modulation
is shielded in the irradiated regions due to enhanced pinning
\cite{banerjee}. Hence, the dark regions in the DMO image (Fig.
1(b)) are the irradiated regions, while the bright background is the
reversible pristine region. Upon increasing $T$, these dark regions
gradually fade until they disappear completely at $H_{irr}^{cd}(T)$,
where the screening current drops to an unobservably low level.

Figure 1(a) shows that upon tilting the field away from the CDs
(changing $\theta$ from $0$ to $90^{\circ}$), the irreversibility
line shifts to lower temperatures. For example, at $H_{z}= 70$ Oe
the irreversibility temperature at $90^{\circ}$ is 2.5~K lower than
at $\theta = 0$. Even at $\theta=90^{\circ}$, $H_{irr}^{cd}(T)$
remains slightly above the melting line of the vortex lattice in the
pristine crystal, $H_{m}^{p}(T)$. This is at odds with the
expectation that at $\theta = 90^{\circ}$ the CDs act as weak point
disorder; then, the melting and irreversibility lines are expected
to lie below $H_{m}^{p}(T)$ \cite{boris}. The angular dependence of
$H_{irr}^{cd}(T)$ is more pronounced at lower temperatures. It
decreases upon increasing $T$ until it disappears completely at
$\sim 82$ K, above which all lines coincide with $H_{m}^{p}(T)$.
This behavior was observed for different samples with $B_{\phi}$ of
20, 30 and 40 G. The relative shift in $H_{irr}^{cd}(T)$ between
$\theta=0$ to $90^{\circ}$ was found to increase with $B_{\phi}$.

To examine the vortex behavior below $H_{irr}^{cd}(T)$, within the
vortex solid phase, we performed conventional MO measurements, in
which a sequence of images is taken while the applied field $H_{a}$
is swept. Figure 1(c)
shows 
flux profiles across the irradiated regions. These have a linear
(Bean-like) shape due to the enhanced pinning in these regions. The
profiles were taken at $H_{z} = 22$ Oe and $T=67$ K, well below
$H_{irr}^{cd}(T)$. At $\theta=90^{\circ}$, the profile is very
shallow. In contrast, at $\theta = 0$, a clear sharp profile is
observed. The critical current, estimated from the flux gradient, is
3.5 times higher at $\theta = 0$ than at $90^{\circ}$.

\begin{figure}[h]
\includegraphics[width=0.32\textwidth]{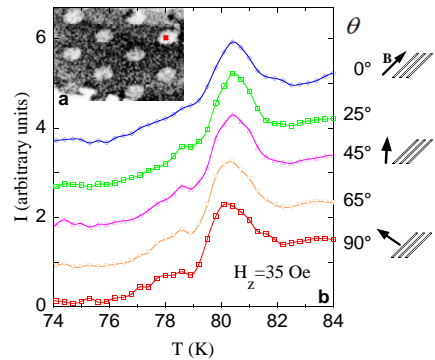}
\caption{\label{fig2}(color online) The melting transition in
crystal B. (a) DMO image of part of the sample at $H_{z}=35$ Oe,
$T=81$ K. (b) Local DMO intensity versus $T$ for different angles
$\theta$. The peak around 80.4 K corresponds to the melting
transition. The curves are vertically displaced for clarity.}
\end{figure}

The angular dependence of the melting transition was studied using
DMO upon temperature sweeps with $T$ modulation \cite{soibel,
banerjee}. The first-order melting transition is characterized by a
step in the equilibrium flux density, and appears as a bright signal
in the DMO images. For example, the bright circular regions in
Fig.~2(a) are the irradiated areas that undergo melting, while the
surrounding pristine regions are already in the liquid phase and
hence appear as a darker background.

Figure 2(b) shows the local DMO intensity, averaged over a region of
$15\times 15$ $\mu$ m$^{2}$ in one of the irradiated regions (red
square in Fig. 2(a)). The melting transition manifests itself as a
peak in the local intensity as $T$ is swept. Each curve corresponds
to a different $\theta$, while $H_{z} = 35$ Oe was kept constant. In
contrast to the irreversibility line that shifts to lower $T$ upon
increasing $\theta$, the melting temperature remains unaffected,
even when $\mathbf{B}$ is perpendicular to the CDs. Moreover, the
transition width, determined from the width of the peak, appears to
be the same for all measured angles. Similar results were obtained
for different $H_{z}$ values. Note that the pristine melting line
was found to have a weak linear dependence on the in-plane field
\cite{schmidt} which cannot be detected within our experimental
resolution for our low in-plane fields.

\begin{figure}[t]
\includegraphics[width=0.41\textwidth]{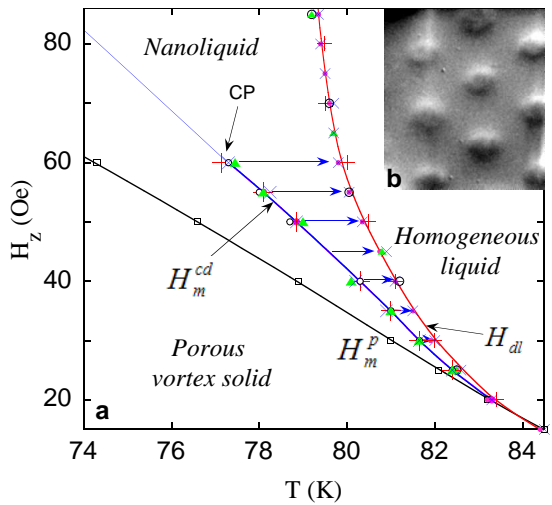}
\caption{\label{fig3}(color online) (a) The melting line
$H_{m}^{cd}(T)$ and the delocalization line $H_{dl}(T)$ in crystal
A, mapped for different $\theta$: $0 (+)$, $25^{\circ}(\circ)$,
$45^{\circ}(\bullet)$, $65^{\circ}(\blacktriangle)$ and
$90^{\circ}(\times)$. (b) DMO self- field image of part of sample A
at $40$ Oe, $75$ K, $20$ mA.}
\end{figure}

Fig.~3(a) gathers the melting lines obtained  at different angles
$\theta$. They all fall on a single line, $H_{m}^{cd}(T)$, that
terminates at a critical point (CP) below which the transition
becomes continuous in nature. The CP
occurs at the same temperature of 77 K at all angles. The width of
the transition, denoted by the horizontal right-pointing arrows, is
broadened on approaching the CP, but remains independent of angle,
as was already shown in
Fig.~2 for $H_{z} = 35$ Oe.  In Fig.~3, we also plot the 
delocalization line $H_{dl}(T)$ \cite {banerjgold}, which separates
the nanoliquid from the homogeneous liquid. $H_{dl}(T)$ was
determined using MO visualization of the self-field induced by a
transport current applied to the sample. DMO image of the self-field
allows one to visualize the transport current density distribution
within the sample. Below $H_{dl}(T)$, the current flow is highly
nonuniform due to enhanced pinning in the irradiated regions
\cite{banerjgold}, and the DMO image shows significant spatial
variations (Fig. 3(b)). These disappear at $H_{dl}$, when the
network of pinned vortices is delocalized, and a homogeneous liquid
is formed. As $H_{m}^{cd}(T)$, the $H_{dl}(T)$ does not depend on
the field angle $\theta$ (Fig. 3(a)). On approaching $T_c$, vortex
pinning by CDs becomes weak and hence all the transition lines
coincide.

The experimental observations show that the dynamic properties of
porous vortex matter are strongly dependent on the angle between
$\mathbf{B}$ and the CDs, while the thermodynamic melting and
delocalization lines are angle-independent. In order to understand
these observations, we have carried molecular dynamics simulations
of a rectangular parallelepiped model system of 36 PV stacks and 200
layers. As shown in \cite{goldcuan,nordborg}, such system size is
sufficient for observing the various phase transitions. Periodic
boundary conditions (PBC) were implemented in all directions. The
intraplane PV interaction was modeled by the repulsion of their
screening currents, while the interplane PV interaction is described
by their mutual electromagnetic attraction. For details see Refs.
[\onlinecite{tyagiyyg,goldcuan}]. We also included the attractive
Josephson interaction between pairs of pancakes in adjacent planes,
but belonging to the same stack \cite{tyagiyyg}. We keep track of
pancakes belonging to a vortex stack and allow for flux cutting and
recombination [\onlinecite{tyagiyyg}]. The average vortex direction
(i.e. ${\rm H_{a}}$) was kept parallel to the $z$ axis, while tilted
CDs of radius $r_r=30$ nm were introduced at random positions. The
interaction between a pancake and a CD in the same plane is given by
an attractive potential \cite{blatter,goldcuan}, which has a long
range tail contribution $\approx -\epsilon_0 d r_r^2/R^2$, due
mainly to the electromagnetic pinning, and a short range flat region
of depth $\approx \epsilon_0 d$ due to the combined effects of
electromagnetic and core pinning. Here,
$\epsilon_0=(\phi_{0}/4\pi\lambda)^{2}$,
$\lambda=\lambda(0)/(1-T/T_c)^{1/2}$ is the penetration depth, $R$
is the distance between a pancake and a CD and $d$ is the layer
separation. The interaction $-\mathbf{B}\cdot \mathbf{H}/(4\pi)$ is
implemented by the PBC, since the electromagnetic interaction
between a tilted stack and the stacks of images, positioned above
(and below) its center of mass, pull the tilted stack in the $z$
direction. Temperature effects are implemented in the simulation by
using a white thermal noise whose variance is proportional to the
temperature.

Figure 4 shows projections of the vortices onto the $xy$ plane (top
view) and the $xz$ plane (side view), with CDs tilted by
$45^{\circ}$ and $80^{\circ}$. The main observation is that, for $B
> B_{\phi}$, the CDs are almost fully decorated by pancakes both in
the solid and nanoliquid phases. This means that the matrix of
pinned PVs is preserved even when the field is tilted away from the
CDs, and as a result the interstitial PVs remain caged in the
surrounding fixed matrix. The interstitial PV stacks are even found
to be tilted to some extent in the direction of CDs, in the solid
phase. Another important observation is that the vortices adopt a
structure featuring kinks between the layers, as shown by the side
views in Figs. 4(c) and 4(d). Kinks of different vortices tend to
align, forming a structure which resembles the Josephson and
Abrikosov crossing lattices \cite{koshelev}  in BSCCO at tilted
field. Here, however, the Josephson vortices (JV) are formed when
tilting the field away from CDs rather than from the $c$-axis. For
example, when $\mathbf{B}$ is parallel to the CDs but at
$45^{\circ}$ with respect to $c$-axis, no JVs are present, while for
$\mathbf{B}\parallel$ $c$-axis JVs segments appear, as shown in
Figs.~4(c) and 4(d).

\begin{figure}
\centering
\includegraphics[width=.44\textwidth]{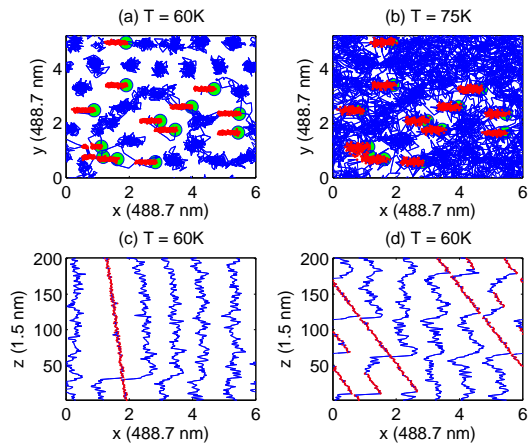}
\caption{(color online) Snapshots of vortex stacks and CDs tilted at
$45^{\circ}$(a, b and c) and $80^{\circ}$(d) for $B=100$ G and
$B^{eff}_{\phi}=35$ G. Pancakes in the same stack are connected.
Free pancakes are in blue, trapped ones in red. Only CDs at the
bottom layer are shown (green). (a)(b) Projection onto the a-b plane
(top view). (a) Nanosolid phase. (b) Nanoliquid phase. (c)
Projection of (a) onto the a-c plane (side view), first row is
shown. (d) Side view with CDs at $80^{\circ}$.} \label{tilted}
\end{figure}

To further interpret the results we consider the energies involved:
The binding energy of a vortex to a CD is of the order of
$\epsilon_0 d$ per layer. For angles less than about $75^{\circ}$
this gain exceeds the electromagnetic energy loss. For an infinitely
long stack tilted at angle $\theta$, the later is given by
$\epsilon_0 d \ln((1+\cos(\theta))/2\cos(\theta))$ per layer
\cite{clem}. In addition, the energy $-\mathbf{B}\cdot
\mathbf{H}/(4\pi)$ tends to align the average vortex orientation
with $\textbf{H}$. The contradicting requirements imposed by the
need to minimize all these energies are optimized by the creation of
kinks. The Josephson energy cost of the kinks is negligible compared
to the total energy of the vortex system in a highly anisotropic
material like BSCCO \cite{bulaevskii}. Indeed, for $\gamma \sim
375$, the total energy measured in the simulations for non tilted
CDs and for CDs tilted at $45^{\circ}$ is the same, in the vicinity
of the melting transition ($T=73$ K to 77 K), within the
simulation's error bars. The comparison was done for the same value
of $B^{eff}_{\phi}=B_{\phi}\cos\theta$ in order to match the
experimental situation where the ratio of vortices to CDs remains
fixed as $\theta$ changes. Our simulations show that the rigid
matrix of pancakes pinned on CDs is preserved up to very large
angles in both the solid and nanoliquid vortex phases. This means
that the enhanced caging potential created by the CDs, that was
found to shift the melting line upwards, \cite {banerjee} is
negligibly affected by increasing the angle between $\mathbf{B}$ and
the CDs. Consequently, the thermodynamic properties, in particular
the melting and delocalization lines, are angle independent. This
was verified in our simulations for $B=100$ G, where the melting and
delocalization transitions occur at 74 K and 76 K respectively, both
for nontilted and tilted CDs, as evidenced by measurements of the
mean square deviation of the flux lines, the amount of their
entanglement and other measurements \cite{goldliu}.

These observations can resolve the apparent discrepancy between the
dynamic and thermodynamic properties. The kinks developed by the
vortices have a negligible effect on the thermodynamics but a large
effect on the dynamics. As demonstrated by our simulations, the
kinks slide along the CDs due to the Lorentz force exerted by
in-plane current, see also \cite{nelson,indenbom}. Consequently,
they transfer PVs between segments, thus producing an effective
transverse drift of the whole pancake stack. Upon increasing
$\theta$, the amount of kinks increases and the flux creep becomes
more effective. This process decreases the critical current and
shifts the irreversibility line to lower temperatures. This is
consistent with \cite {drost}, which concluded that CDs' pinning
anisotropy commonly measured in BSCCO \cite{irreversible} must be a
dynamic effect.

In summary, the porous vortex matter in BSCCO is shown to preserve
its thermodynamic properties even when the field is tilted away from
the CDs. While the irreversibility line shifts to lower temperatures
upon tilting the field, the thermodynamic melting and delocalization
lines remain constant. Numerical simulation shows that increasing
the angle between the field and the CDs leads to formation of weakly
pinned vortex kinks, while preserving the basic structure of a rigid
matrix of pancake stacks residing along the CDs with nanocrystals of
interstitial vortices embedded within the pores of the matrix.

We thank S. Goldberg and Ady Stern for stimulating discussions, D.
Linsky for technical help. This work was supported by the ISF Center
of Excellence, German-Israeli Foundation (GIF), US-Israel BSF and
Grant-in-aid from the Ministry of Education, Culture, Sports,
Science, and Technology, Japan. The work of YYG and JTL was
supported by the US Department of Energy (DOE-DE-FG02-98ER45686),
NERSC program and the Pittsburgh Supercomputing center. YYG thanks
the Weston Visiting Professors program of the WIS for support.


\begin{thebibliography}{99}
\bibitem{angle} Tilting $\mathbf{B}$ by $\pm45^{\circ}$ requires tilting $H_{a}$  by
$\pm40^{\circ}$ (found experimentally) since $H_{z}$ is screened
more efficiently by in-plane equilibrium shielding currents,
compared to $H_{ab}$.
\end{thebibliography}
\end{document}